\documentclass[11pt,twoside]{article}
\usepackage{geometry}
\usepackage{moreverb,url}
\usepackage[normalem]{ulem}   
\usepackage{graphicx}
\usepackage[colorlinks,bookmarksopen,bookmarksnumbered,citecolor=red,urlcolor=red]{hyperref}
\usepackage{booktabs}
\usepackage{multirow}
\usepackage{footnote}
\usepackage{float}
\usepackage{multirow}
\usepackage{rotating}
\usepackage{subcaption}
\makesavenoteenv{tabular}

\newcommand\BibTeX{{\rmfamily B\kern-.05em \textsc{i\kern-.025em b}\kern-.08em
T\kern-.1667em\lower.7ex\hbox{E}\kern-.125emX}}

\begin{document}

\title{\Large Survival analysis for AdVerse events with VarYing follow-up times (SAVVY) ---
  estimation of adverse event risks}
  
\author{Regina Stegherr$^{1}$, Claudia Schmoor$^{2}$, Jan Beyersmann$^{1,*}$, Kaspar Rufibach$^{3}$, Valentine Jehl$^{4}$,\\ Andreas Br\"uckner$^{4}$, Lewin Eisele$^{5}$, Thomas K\"unzel$^{3}$, Katrin Kupas$^{6}$, Frank Langer$^{7}$,\\ Friedhelm Leverkus$^{8}$, Anja Loos$^{9}$, Christiane Norenberg$^{10}$, Florian Voss$^{11}$ and Tim Friede$^{12}$}
\date{August 13, 2020}
\maketitle
\noindent${}^{1}$ Institute of Statistics, Ulm University, Ulm, Germany\\
\noindent${}^{2}$ Clinical Trials Unit, Faculty of Medicine and Medical Center, University of Freiburg, Freiburg im Breisgau, Germany\\
\noindent${}^{3}$ F. Hoffmann-La Roche, Basel, Switzerland\\
\noindent${}^{4}$ Novartis Pharma AG, Novartis Pharma AG, Basel, Switzerland\\
\noindent${}^{5}$ Janssen-Cilag GmbH, Neuss, Germany\\
\noindent${}^{6}$ Bristol-Myers-Squibb GmbH \& Co. KGaA, München, Germany\\
\noindent${}^{7}$ Lilly Deutschland GmbH, Bad Homburg, Germany\\
\noindent${}^{8}$ Pfizer, Berlin, Germany\\
\noindent${}^{9}$ Merck KGaA, Darmstadt, Germany\\
\noindent${}^{10}$ Bayer AG, Wuppertal, Germany\\
\noindent${}^{11}$ Boehringer Ingelheim Pharma GmbH \& Co. KG, Ingelheim, Germany\\
\noindent${}^{12}$ Department of Medical Statistics, University Medical Center G\"ottingen, G\"ottingen, Germany\\

\noindent${}^{*}$ {Corresponding author: Jan Beyersmann, jan.beyersmann@uni-ulm.de}

\textbf{Abstract}\\
  \textbf{Background:} The SAVVY project aims to improve the analyses of
  adverse event (AE) data in clinical trials through the use of survival
  techniques appropriately dealing with varying follow-up times and competing
  events. Although statistical methodologies have advanced, in AE analyses
  often the incidence proportion, the incidence density, or a non-parametric
  Kaplan-Meier estimator are used, which either ignore censoring or competing
  events.  In an empirical study including randomized clinical trials from
  several sponsor organisations, these potential sources of bias are
  investigated. The main purpose of the empirical study is to compare the
  estimators that are typically used in AE analysis to the non-parametric
  Aalen-Johansen estimator as the gold-standard. The present paper reports on
  one-sample findings, while a companion paper considers consequences when
  comparing safety between treatment groups. \\
  \textbf{Methods:} Estimators are compared with descriptive statistics, graphical displays and
  with a more formal assessment using a random effects meta-analysis. The influence of different factors on the size of the bias is investigated in a
  meta-regression. Comparisons are conducted at the maximum follow-up time and at earlier evaluation time points. Competing events
  definition does not only include death before AE but also end of follow-up for AEs due to events possibly related to the disease course or safety of the treatment.\\
  \textbf{Results:} Ten sponsor organisations provided 17
  clinical trials including 186 types of investigated AEs. The one minus Kaplan-Meier
  estimator was on average about 1.2-fold larger than the Aalen-Johansen
  estimator and the probability transform of the incidence density ignoring competing events
  overestimated the AE probability even more. Leading forces influencing bias
  were the amount of censoring and of competing events. The presence of
  many competing events in our study decreased the amount of censoring. As a
  consequence, the average bias using the incidence proportion was less than
  5\%. Assuming constant hazards using incidence densities was hardly an issue
  provided that competing events were accounted for.\\
  \textbf{Conclusions:} Both the choice of the estimator of the cumulative AE
  probability and a careful definition of competing events are crucial. There
  is an urgent need to improve the guidelines of reporting risks of AEs so that the
  Kaplan-Meier estimator and the incidence proportion are finally replaced by
  the Aalen-Johansen estimator with an appropriate definition of competing
  events.\vspace{2cm}

\textbf{Keywords:} Aalen-Johansen estimator, adverse events, competing events, drug safety, incidence proportion, incidence density, Kaplan-Meier estimator

\section{Background}
Time-to-event or survival endpoints are common in clinical research
\cite{Hort:Switz:2005,sato2017statistical}. The observation of the event times
is typically incomplete as a consequence of censoring, and the statistical
analysis, therefore, requires specialized techniques. This requirement holds for
both the evaluation of efficacy and safety. An important aim of the latter is the estimation of the probability of an adverse event (AE) of a specific type within a specific time interval, which, in a time-to-first-event analysis, is often done by the incidence
proportion, i.e., the number of patients with an observed AE (of a certain
type) in a specific time period divided by group size, or the (exposure adjusted) incidence density,
which divides by cumulative patient-time at risk. The worry is that the
incidence proportion underestimates the cumulative AE probability because it
does not account for censoring \cite{oneill_1987,allignol2016,bender_2016,
  unkel2018}. One minus a Kaplan-Meier estimator counting AEs as the event
would account for censoring, but not for competing events (CEs) such as death
without prior AE. When CEs are present, Kaplan-Meier is commonly
used \cite{schumacher2016competing,van2016competing} but bound to overestimate
the cumulative AE probability as the methodology implicitly assumes that every
patient experiences the AE under consideration, possibly after a CE such as death.
The incidence density also accounts for censoring but does not estimate a
probability. Rather, it estimates the AE hazard assuming it to be
time-constant. However, this assumption is not realistic for many drug-related adverse events\cite{Krae:even:2009, bender_2019}. The interpretation of the incidence density as an estimator of a hazard is challenging, but the incidence density may be transformed
onto the probability scale; typically, such transformations do not consider
CEs \cite{cummings19:_analy_incid_rates}, although extensions are
available \cite{bonofiglio2015meta}.

The concerns above are qualitative. However, the amount of bias, comparing,
e.g., the incidence proportion or one minus Kaplan-Meier with the
non-parametric gold standard, the Aalen-Johansen estimator \cite{allignol2016}
accounting for both CEs and censoring will depend on the
specific trial setting. In particular, the relative frequencies of observed
AEs, observed CEs and observed censorings add up to 100\% at any
point in time. The latter two are leading forces influencing bias, and, e.g.,
the presence of many CEs in a time-to-first-event analysis will
impact the amount of censoring.

The SAVVY project group (Survival analysis for AdVerse events with Varying
follow-up times) is a collaborative effort from academia and pharmaceutical industry with the
aim to improve the analyses of AE data in clinical trials through the use of
survival techniques that account for varying follow-up times, censoring
and CEs. Here, we report one-sample results from an empirical
study of an opportunistic sample of randomized clinical trials from several
sponsor companies. The aim is to illustrate possible biases when quantifying
absolute AE risk in single samples including categorization into AE frequency
categories. Results when comparing safety between treatment groups in the
two-sample case are in a companion paper\cite{twosample}.  Individual
trial data analyses were run within the sponsor organisations using SAS and R
software provided by the academic project group members. Only aggregated data
necessary for meta-analyses were shared and meta-analyses were run centrally
at the academic institutions.

\section{Methods}
A detailed Statistical Analysis Plan is available elsewhere
\cite{stegherr2019survival}. Here, we briefly summarize one-sample estimators
and methods of meta-analysis. Properties and estimands of the estimators are discussed elsewhere\cite{stegherr2019survival,unkel2018}. We describe in more detail the definition of CEs which has an immediate consequence on the estimation procedures.

\subsection{One-sample estimators}
We will consider the following estimators of the cumulative AE probability or
`AE risk' in a time-to-first-event analysis. Since both probabilities and the amount of censoring
\cite{pocock2002survival} are time-dependent, we will allow for different evaluation times called~$\tau$. These evaluation times either imposed no restriction,
i.e., evaluated the estimators until the maximum follow-up time, or considered
the minimum of quantiles of observed times in the two treatment groups; the quantiles were
100\%, 90\%, 60\% and 30\%. We will report results from `Arm E', denoting the experimental treatment groups. 
The incidence proportion is
\begin{equation}
  \label{eq:1}
  IP_{\rm E}(\tau) = \frac{\mbox{no. of patients w.\ observed AE on
      $[0, \tau]$ in E}}{n_{\rm E}},
\end{equation}
where~$n_{\rm E}$ denotes sample size in group E. 
This estimator will be called \emph{incidence proportion} in the following.

The AE incidence density is
\begin{equation}
  \label{eq:2}
  ID_{\rm E}(\tau) = \frac{\mbox{no. of patients w.\  observed AE on
      $[0, \tau]$ in E}}{\mbox{patient-time at risk in E
      restricted by $\tau$}}
\end{equation}
which we transform onto the probability scale using
\begin{equation}
  \label{eq:3}
  1 - \exp\left(- ID_{\rm E}(\tau) \cdot \tau\right),
\end{equation}
called \emph{probability transform incidence density ignoring CE} in the following.
The \emph{one minus Kaplan-Meier} 
estimator only codes observed AEs as
an event and censors anything else on~$[0, \tau]$.

An incidence densities analysis accounting for CEs uses the competing incidence density
\begin{equation}
  \label{eq:4}
  \overline{ID}_{\rm E}(\tau) = \frac{\mbox{no. of patients w.\ 
      observed CE on $[0, \tau]$ in E}}{\mbox{patient-time at risk in E restricted by $\tau$}}
\end{equation}
such that we get the following AE-probability estimator
\begin{equation}
  \label{eq:5}
  \frac{ID_{\rm E}(\tau)}{ID_{\rm E}(\tau) + \overline{ID}_{\rm
      E}(\tau)} \left(1-\exp(-\tau\cdot[ID_{\rm E}(\tau) + \overline{ID}_{\rm
      E}(\tau)])\right),
\end{equation}
called \emph{probability transform incidence density accounting for CE} in the following.
Finally, the \emph{Aalen-Johansen} estimator generalizes (\ref{eq:5})
to a fully non-parametric procedure and decomposes the usual one minus
Kaplan-Meier estimator of the time-to-\emph{any}-first-event (AE or competing)
into estimators of the cumulative AE probability plus the cumulative CE
probability \cite{allignol2016}. 

\subsection{Definition of competing events}
The definition of events as `competing' is essential to both the
Aalen-Johansen estimator and the competing incidence density. CEs
(or `competing risks') are events that preclude the occurrence or recording of the AE under
consideration in a time-to-first-event analysis. One important competing event
is death before AE. In addition, any event that would both be viewed from a patient perspective as an event of his/her course of disease or treatment and would stop the recording of the interesting AE will be viewed as a CE. To illustrate, premature discontinuation of study treatment which leads to end of AE recording will be handled as a CE\cite{jbcs2019}. Consequently, possibly disease- or safety-related
loss to follow-up, withdrawal of consent and discontinuation is handled as a
competing event as this is typically related to an event associated with the disease course or therapy.

In order to investigate the impact of the definition of CEs, we
also investigated a `death only' scenario, which only treated death before AE
as competing, but not the other CEs. This estimator will be called \emph{Aalen-Johansen (death only)} 
in the following. 

The data generation mechanism underlying the clinical trials is based on the hazard of the AE, the hazard of the CE, and the distribution of the censoring times, where the hazards are not restricted to be constant\cite{stegherr2019survival}. But not all estimators suggested for analysing AEs can adequately deal with all three processes. Table~\ref{tab:14top} gives an overview whether the estimators account for the three sources of bias, i.e., censoring, no constant hazards, and CEs.  The incidence proportion ignores CEs and censoring in the analysis in the same way as the respective patients are counted in the denominator as if they had been followed for the entire study period. This is a proper handling of the CEs as it correctly takes into account that an AE cannot occur after the patient had experienced a CE. It is an improper handling of censoring as it incorrectly implies that an AE could have been observed over the entire follow-up period, which is not true due to censoring.  

The Aalen-Johansen estimator is the only estimator that is able to deal with all three potential sources of bias and is therefore considered the gold standard estimator and
will serve as a benchmark for comparison of results. In the
following, we will use the term bias for deviations of the
estimators from this benchmark estimator and not for the
difference to the true value. This is considered appropriate as the differences of
the estimators to the Aalen-Johansen estimator converge in
probability to the asymptotic bias.

\begin{table}
\small\sf\centering
\caption{Overview whether the estimators deal with the possible sources of bias.\label{tab:14top}}
 \begin{tabular}{lccc}
 \toprule
&Accounts for & Makes no constant & Accounts for \\
&censoring & hazard assumption & CEs\\
\midrule
Incidence proportion & No & Yes & Yes\\
Probability transform incidence density & Yes & No (AE Hazard) & No\\
{\hspace{0.2cm}} ignoring CEs & & & \\
1-Kaplan-Meier & Yes & Yes & No\\
Probability transform incidence density  & Yes & No (AE and CE Hazard) & Yes\\
{\hspace{0.2cm}}accounting for CEs & & & \\
death only Aalen-Johansen estimator & Yes & Yes & Yes (Death only)\\
gold-standard Aalen-Johansen estimator & Yes & Yes & Yes \\
\bottomrule
\end{tabular}
\end{table}

\subsection{AE frequency categories}
According to the European Commission's guideline on summary of product characteristics
(SmPC)\cite{smpc} and based on the recommendations of the CIOMS Working Groups III and V\cite{AEcat} the frequency categories of AE risk in the most representative exposure period are respectively classified as `very rare', `rare', `uncommon', `common'  and `very common' when found to be $<$ 0.01\%, $<$ 0.1\%, $<$ 1\%, $<$ 10\%, $\ge$ 10\%. 
Frequency categories obtained with the different estimators will be compared to frequency categories obtained with the gold-standard Aalen-Johansen estimator.

\subsection{Random effects meta-analysis and meta-regression}
In the meta-analysis and meta-regression, the ratios of the AE probability estimates obtained with the different estimators divided by the AE probability obtained with the gold-standard Aalen-Johansen estimator are considered on the log-scale. The standard errors of these log-ratios are calculated with a bootstrap to account for within trial dependencies. Then, a normal-normal hierarchical model is fitted and the exponential of the resulting estimate can be interpreted as the average ratio of the two estimators. 

In a meta-regression it is further investigated which variables impact this average ratio. Therefore, the proportion of censoring, the evaluation time point $\tau$, i.e., the maximal time to event in years (AE, CE or censoring) observed under the given evaluation time, and the size of the AE probability estimated by the gold-standard Aalen-Johansen are included as covariates in a univariable and a multivariable meta-regression. The covariates are centered in the meta-regression.

\section{Results} 

\subsection{Description of the data}Ten organisations provided 17 trials including 186 types of AEs (median $8$; interquartile range $[3, 9]$). Twelve (71.6\% out of 17) trials were from
oncology, nine (52.9\%) were actively controlled and eight (47.1\%) were
placebo controlled. The trials included between 200 and 7171 patients (median: 443; interquartile range $[411, 1134]$). For the comparison of the AE probabilities we focus on the experimental treatment group. The corresponding results of the control group are displayed in the companion paper\cite{twosample}. Median follow-up of the treatment group was 927 days (interquartile range $[449, 1380]$). In the experimental treatment group, the median of the calculated gold-standard Aalen-Johansen estimators was 0.092 (minimum 0 and maximum 0.961).

\begin{figure*}
\centering
\includegraphics{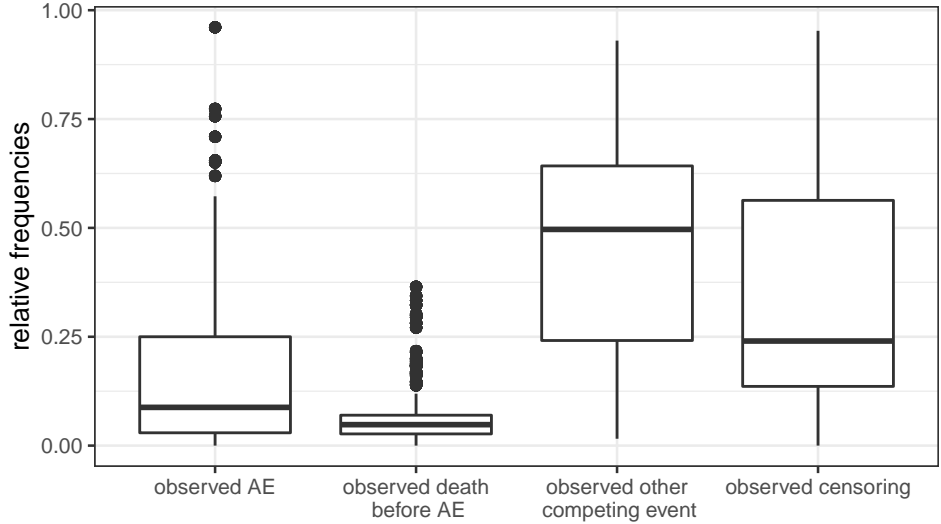}
\caption{Relative frequencies of observed events\label{fig:tab:A1}}
\end{figure*}

Figure~\ref{fig:tab:A1} displays for the 186 types of AEs boxplots of the observed relative frequencies, i.e., the number of patients with a specific type of event divided by the total number of patients, namely of `observed AE', `observed death before AE', `observed other competing events', and
`observed censoring' for the maximal follow-up time.

The figure illustrates a smaller amount of observed censoring compared to
observed other, i.e., non-death CEs. That is, AE recording often ended due to death or other CEs such as treatment discontinuation preventing censoring of the time to AE. There are also much less death events than other CEs.

\subsection{Comparison of AE probability estimators}

\begin{figure*}
\centering
\includegraphics{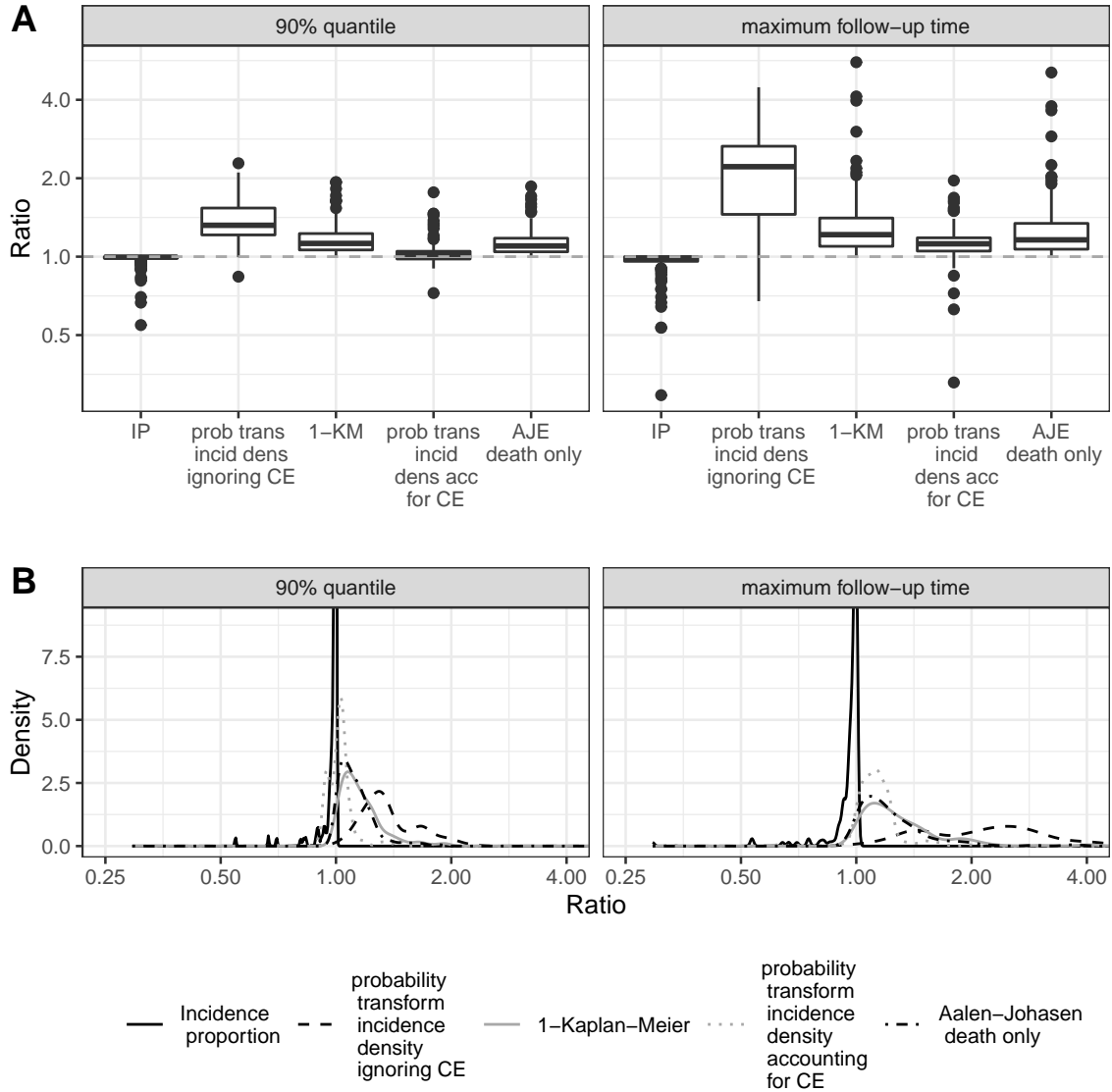}
\caption{Panel A: Accepting the all event definition of competing events as gold-standard, the ratios of one-sample estimator divided by gold-standard Aalen-Johansen estimator are displayed. Two different evaluation times are displayed. The left boxplots are the results for the estimators being evaluated at the 90\% quantile and the right boxplots are the results of the evaluation time with no restriction, i.e., at the end of follow-up. The following abbreviations are used for the estimators: incidence proportion (IP), probability transform of the incidence density ignoring CE (prob trans incid dens ignoring CE), one minus Kaplan-Meier (1-KM), probability transform of the incidence density accounting for CE (prob trans incid dens acc for CE), death only Aalen-Johansen estimator (AJE death only). Panel B: Plots of the kernel density estimates of the ratios of the AE probability of the estimators divided by the gold-standard Aalen-Johansen estimator.\label{fig:tab:15}}
\end{figure*}

Panel A of Figure~\ref{fig:tab:15} shows box plots of the ratio of the one-sample estimators defined earlier 
divided by the gold-standard Aalen-Johansen estimator for the maximum follow-up time and one earlier evaluation time chosen as to the 90\% quantile. As the incidence proportion
implicitly accounts for CEs (but not for censoring) as explained above, the small amount of censoring which is a consequence of the high amount of other CEs explains why the incidence proportion and the Aalen-Johansen estimator are of similar size in many situations. But it has to be emphasized that in extreme cases an underestimation of up to seventy percent was present. 

Both one minus Kaplan-Meier and the probability transform incidence density ignoring CE overestimate the AE probability, and this is also true for the Aalen-Johansen estimator that only considers death before AE as competing. Interestingly, the probability transform incidence density ignoring CE appears to be worst, while the probability transform incidence density accounting for CE performs much better
than the other three procedures which are clearly biased resulting in extreme overestimation in many situations, up to a factor of five. 
These biases become less pronounced when looking at earlier evaluation times which prevent CEs and censoring after the respective end of evaluation time to enter calculations. 

\subsection{Impact on frequency categories}
The impact on frequency categories is illustrated in
Table~\ref{tab:tab:A5:bis:A9}, where we have exemplarily chosen the
maximum follow-up time as most representative exposure period.
\begin{table}
\small\sf\centering
\caption{The impact of the choice of one-sample estimator on AE frequency categories for the maximal follow-up time. Deviations from the Aalen-Johansen estimator are the non-diagonal entries. The first rows consider the gold-standard Aalen-Johansen estimator and the last five rows the comparison of the incidence proportion and the Aalen-Johansen (death only) estimator. Diagonal entries are set in bold face. Non-diagonal zeros are omitted from the display.\label{tab:tab:A5:bis:A9}}
 \begin{tabular}{@{}c@{}c@{}c@{}c@{}c@{}c@{}clccccc}
 \toprule
&&& &&&&& \multicolumn{5}{c}{gold-standard Aalen-Johansen}\\
&&&&&&& &very rare & rare & uncommon & common & very common\\
\midrule
& &&\multirow{5}*{\rotatebox{90}{incidence}} & \multirow{5}*{\rotatebox{90}{proportion}} 
&&&very rare &\textbf{6} && & &\\
&&&&&&&rare & & \textbf{0}&& &\\
&&&&&&&uncommon & & & \textbf{6}& &\\
&&&&&&&common & && &\textbf{86} &2\\
&&&&&&&very common && & & &\textbf{86}\\
\midrule
 &\multirow{5}*{\rotatebox{90}{probability}} &\multirow{5}*{\rotatebox{90}{transform}} & \multirow{5}*{\rotatebox{90}{incidence}}& \multirow{5}*{\rotatebox{90}{density}} & \multirow{5}*{\rotatebox{90}{ignoring CE}} 
&&very rare &\textbf{6} & & & &\\
&&&&&&&rare & & \textbf{0}&& &\\
&&&&&&&uncommon & & & \textbf{3}& &\\
&&&&&&&common & && 3&\textbf{51} &\\
&&&&&&&very common && & &35 &\textbf{88}\\
\midrule
 &&&   \multirow{5}*{\rotatebox{90}{1-Kaplan-}}& \multirow{5}*{\rotatebox{90}{Meier}} 
&&&very rare &\textbf{6} & & &&\\
&&&&&&&rare & & \textbf{0}&& &\\
&&&&&&&uncommon & & & \textbf{4}& &\\
&&&&&&&common & && 2&\textbf{72}&\\
&&&&&&&very common & &  & &14&\textbf{88}\\
\midrule
 \multirow{5}*{\rotatebox{90}{probability}} &\multirow{5}*{\rotatebox{90}{transform}} & \multirow{5}*{\rotatebox{90}{incidence}}& \multirow{5}*{\rotatebox{90}{density}} & \multirow{5}*{\rotatebox{90}{accounting}}& \multirow{5}*{\rotatebox{90}{for CE}} 
&&very rare &\textbf{6} & & & &\\
&&&&&&&rare & & \textbf{0}&& &\\
&&&&&&&uncommon & & & \textbf{4}& &\\
&&&&&&&common & && 2&\textbf{79}&1\\
&&&&&&&very common & & & &7&\textbf{87}\\
\midrule
& & \multirow{5}*{\rotatebox{90}{Aalen-}}& \multirow{5}*{\rotatebox{90}{Johansen}} & \multirow{5}*{\rotatebox{90}{death only}}& 
&&very rare &\textbf{6} && & &\\
&&&&&&&rare & & \textbf{0}&& &\\
&&&&&&&uncommon & & & \textbf{5}& &\\
&&&&&&&common & && 1&\textbf{73}&\\
&&&&&&&very common & & &&13&\textbf{88}\\
\midrule\midrule
&&& &&&&& \multicolumn{5}{c}{Aalen-Johansen death only}\\
&&&&&&& &very rare & rare & uncommon & common & very common\\
\midrule
& &&\multirow{5}*{\rotatebox{90}{incidence}} & \multirow{5}*{\rotatebox{90}{proportion}} 
&&&very rare &\textbf{6} & & & &\\
&&&&&&&rare & & \textbf{0}&& &\\
&&&&&&&uncommon & & & \textbf{5}& 1&\\
&&&&&&&common & &&&\textbf{73} &15\\
&&&&&&&very common && & & &\textbf{86}\\
\bottomrule
\end{tabular}
\end{table}
Some switches to neighboring categories are detected. The probability transform of the incidence density ignoring CEs derives a higher AE frequency category for 38 types of AEs, and the one minus Kaplan-Meier estimator for 16 types of AEs. The probability transform of the incidence density accounting for CE obtains a higher category for nine types of AEs but also a lower category for one type of AE. Here, the definition of the competing event is again of importance. The death only Aalen-Johansen estimator categorizes 14 types of AEs to a higher category than the gold-standard Aalen-Johansen estimator. The incidence proportion derives only two times a different AE frequency category than the gold-standard Aalen-Johansen estimator. The good performance of the incidence proportion is closely connected to the CE definition, i.e. the maturity of data at the time of the analysis. If in the comparison to the incidence proportion the Aalen-Johansen (death only) is used instead of the gold-standard, the category common instead of very common is obtained for 15 types of AEs and one type of AE is categorized to uncommon using the incidence proportion but to common using the Aalen-Johansen that only considers death as a competing event estimator (see last five rows of Table~\ref{tab:tab:A5:bis:A9}).

\subsection{Random effects meta-analysis}
In a meta-analysis of the log-ratio of the incidence proportion divided by the
Aalen-Johansen estimator evaluated at the maximum follow-up time, the average
ratio was found to be~$0.972$ with a 95\%-confidence interval
of~$[0.965, 0.980]$. The respective result for the probability transform incidence density ignoring CE 
was $2.097\ [1.994, 2.205]$ and for one minus Kaplan-Meier was $1.214\ [1.184, 1.245]$. Accounting for competing
risks in an incidence densities-analysis (probability transform incidence density accounting for CE) gave a result of
$1.130\ [1.112, 1.150]$, while the Aalen-Johansen (death only) estimator lead
to an average of $1.170\ [1.145, 1.195]$.
These results confirm the visual impression gathered from the boxplots in Panel A of 
Figure~\ref{fig:tab:15}, but we note that Panel A of Figure~\ref{fig:tab:15} also
displays biases in individual trials which are much larger than the meta-analytical averages.
\subsection{Random effects meta-regression}
The influence of different factors on the size of the bias was investigated in univariable and multivariable meta-regression. The percentage of censoring, the size of the AE probability estimated by the gold-standard Aalen-Johansen, and the evaluation time point were considered and included as covariates in the meta-regression models. In Tables~\ref{tab:tab:16} 
results are exemplarily displayed when evaluating estimators using the maximum follow-up time as evaluation time. 
\begin{sidewaystable}
\small\sf\centering
\caption{Univariable and multivariable meta-regression. Average risk ratio and multiplicative change by 10\% increase in censoring, 10\% increase in CEs, one additional year of observation or a 0.1 greater AE probabiltiy. Thereby, the size of the AE probability is estimated by the gold-standard Aalen-Johansen estimator.\label{tab:tab:16}}
    \begin{tabular}{llccccc}
 \toprule
&& & probability transform& & probability transform & \\
&&incidence &incidence density&1-Kaplan-Meier & incidence density & Aalen-Johansen\\
&&proportion & ignoring CE && accounting for CE &death only\\
\midrule
 \multicolumn{2}{l}{\textbf{Univariable meta-regression}}&&&&\\
\% censoring &average risk ratio& 0.974 [0.964; 0.983] & 2.308 [2.217; 2.403] & 1.257 [1.226; 1.288] & 1.101 [1.086; 1.116] & 1.201 [1.175; 1.228]\\
&10\% increase& 0.999 [0.996; 1.002] &  0.916 [0.903; 0.929] & 0.973 [0.965; 0.980]& 1.026 [1.021; 1.031]&0.979 [0.972; 0.986]\\
\midrule
\%CEs &average risk ratio&0.976 [0.969; 0.984] &2.191 [2.141; 2.243] & 1.240 [1.214; 1.267] &1.124 [1.109; 1.140] & 1.190 [1.168; 1.213] \\
&10\% increase & 1.003 [1.000; 1.006] & 1.127 [1.117; 1.138] & 1.036 [1.028; 1.045] &0.977 [0.971; 0.982]  & 1.029 [1.021; 1.036]\\
\midrule
size of AE &average risk ratio& 0.973 [0.966; 0.980] & 2.105 [2.005; 2.210] & 1.215 [1.185; 1.246] & 1.131 [1.112; 1.150] & 1.171 [1.146; 1.197]\\
probability&increase of 0.1& 0.996 [0.992; 1.000] & 0.954 [0.930; 0.980] & 0.995 [0.982; 1.008]& 0.993 [0.984; 1.003]& 0.993 [0.982; 1.004]\\
\midrule
evaluation&average risk ratio & 0.972 [0.964; 0.980] & 2.094 [1.994; 2.199] & 1.214 [1.184; 1.244] & 1.131 [1.112; 1.150] & 1.170 [1.145; 1.195]\\ 
 time&one additional year & 0.993 [0.987; 1.000] & 1.054 [1.021; 1.087] & 1.015 [0.999; 1.033] & 0.996 [0.986; 1.007] & 1.013 [0.998; 1.027]\\
\midrule\midrule
 \multicolumn{2}{l}{\textbf{Multivariable meta-regression}}&&&&\\
 \multicolumn{2}{l}{average risk ratio} & 0.976 [0.966; 0.985] & 2.407 [2.348; 2.468] & 1.277 [1.246; 1.308] & 1.097 [1.082; 1.113] & 1.218 [1.192; 1.245]\\
\multicolumn{2}{l}{\%censoring 10\% increase} &  0.997 [0.994; 1.000] & 0.890 [0.882; 0.899] & 0.965 [0.957; 0.973] & 1.028 [1.023; 1.034]&0.972 [0.965; 0.979]\\
 \multicolumn{2}{l}{size of AE probability increase of 0.1}&0.995 [0.991; 0.999]& 0.893 [0.882; 0.904] & 0.972 [0.961; 0.983] & 1.008 [1.000; 1.016] & 0.975 [0.965; 0.985]\\
 \multicolumn{2}{l}{evaluation time one additional year}&0.994 [0.988; 1.000] &1.036 [1.021; 1.051] & 1.014 [1.000; 1.027] & 1.003 [0.995; 1.011] & 1.011 [0.999; 1.024]\\
\bottomrule
\end{tabular}
\end{sidewaystable}

Covariates were centered, i.e., the row `average risk ratio' contains the average ratio of the estimator of interest and the Aalen-Johansen estimator if the covariate takes its mean. Those means were 31.5\% censoring, 52.6\% competing events, 971 days maximum follow-up time, and a size of the AE probability estimated by the Aalen-Johansen estimator of 0.165. For example, for the comparison of the incidence proportion and the Aalen-Johansen estimator the estimated average ratio of the two estimators in a trial with 31.5\% censoring is 0.974. Furthermore, in a trial with 10\% more censoring the estimated average ratio is increased by the factor 0.999 but the unit value is contained in the corresponding confidence interval. So, the amount of underestimation 
by the incidence proportion which does not account for censoring slightly increases 
with an increasing amount of censoring.
Considering the estimators that either do not (probability transform incidence density ignoring CE, 
one minus Kaplan-Meier) or only partially (Aalen-Johansen (death only)) account for CE, one finds that both a higher amount of censoring and a higher AE probability decrease the amount of overestimation. The explanation goes hand in hand with the increased average ratios for higher amounts of CEs as these estimators do account for censoring, and increased censoring will, in general, lead to a smaller amount of observed competing events. Likewise, a higher AE probability will, 
in general, lead to a smaller probability of CEs.

These results are confirmed by the multivariable meta-regression.
The amount of CEs is not included in the multivariable meta-regression as there is a strong dependence with the amount of censoring and the size of the AE probability estimated by the gold-standard Aalen-Johansen estimator.

\subsection{Variability}
Even though on average the incidence proportion does well in this sample of selected AEs the possible variability must not be neglected. 


Considering the plots of the kernel density estimates of the ratios of the different estimators of the AE probability in Panel B of Figure~\ref{fig:tab:15}, the ratio of incidence proportion and the gold standard is most often close to one. But there are also peaks of the estimated kernel density at smaller ratios indicating that the estimators are not always comparable. For the ratio of the probability transform of the incidence density accounting for CEs and the gold standard most values are slightly larger than one at the maximum follow-up time. At the earlier follow-up time according to the 90\% quantile the peak is closer to one with less variability present. The ratios of the one minus Kaplan-Meier and death only Aalen-Johansen estimator to the gold standard have few values close to one. For the majority of AE types these two estimators largely overestimate the AE probability. Both plots illustrate pronounced variability for probability transform of the incidence density ignoring CE.

\subsection{Exemplary results from single trials }
A closer look is taken at single AE types in trials for which extreme under- or overestimation is present, i.e. extreme values in the right panel boxplots in Figure~\ref{fig:tab:15}. For example, the largest underestimation of the incidence proportion is for an AE which is only observed for three out of 274 patients. This corresponds to an incidence proportion of 0.011. However, an Aalen-Johansen estimate of 0.037 is obtained. This corresponds to a ratio of 0.294 with a 95\% confidence interval of [0.084; 1.025], where the confidence interval has been obtained using the bootstrap. As 27.0\% of the observations for this type of AE are censored, the amount of censoring is below the mean censoring rate of all types of AEs. Moreover, for this type of AE 17 deaths (6.2\%) and 180 other CEs (65.7\%) are observed. This type of AE does not only contribute the largest underestimation of the incidence proportion but also of the probability density of the incidence density accounting for CEs for which an estimate of 0.012 is obtained (ratio of 0.329 with 95\% CI [0.094; 1.148]). Furthermore, for this type of AE the largest overestimation of the one minus Kaplan-Meier estimator (estimate of 0.208 and ratio of 5.575 [1.813; 17.147]) and the Aalen-Johansen (death only) estimator (estimate of 0.190 and ratio of 5.090 [1.815; 14.276]) is calculated. These impressive ratios
are partly due to the small value of the gold-standard Aalen-Johansen estimate, but we stress that also the difference between one minus Kaplan-Meier and the gold standard is quite pronounced (0.208 vs. 0.037). 

In another extreme example with a higher AE probability the obtained incidence proportion is 0.059 and the Aalen-Johansen estimate is 0.109 (ratio 0.534 [0.529; 0.540]). For this type of AE many censored observations are present (63.3\% of 752 patients). Moreover, 44 AEs are observed, 137 deaths (18.2\%), and 95 other CEs (12.6\%). Here, due to the high amount of censoring one can expect in advance the incidence proportion not doing well.

\subsection{Role of Censoring}
  
\begin{figure*}
\centering
\includegraphics{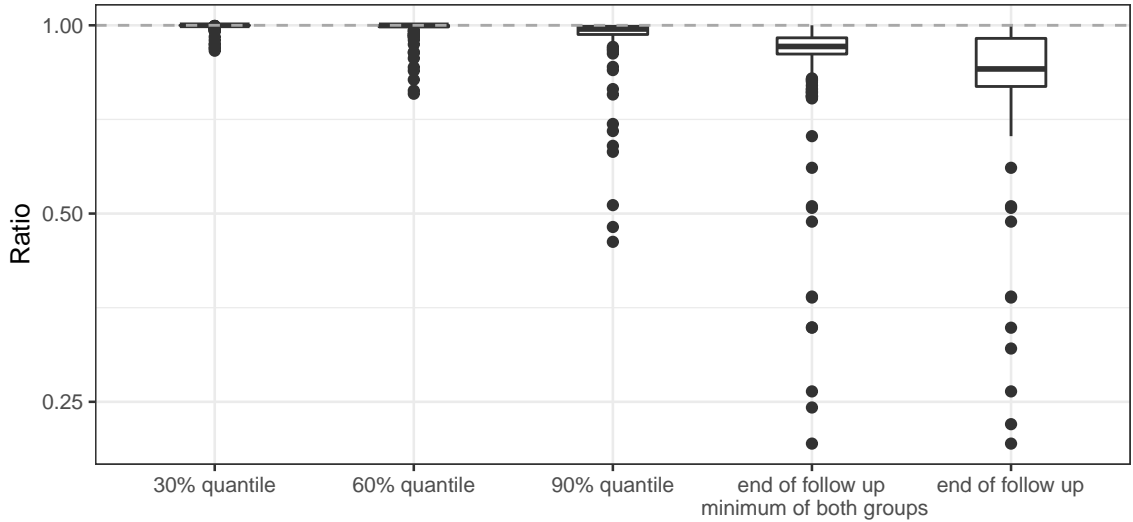}
\caption{Ratios of incidence proportion of the composite endpoint combining AE and CE divided by composite 1-Kaplan-Meier estimator\label{fig:composite}}
\end{figure*}

To explicitly investigate the role of censoring without the methodological complication of CEs the composite endpoint combining AEs and CEs is considered, which results in a single endpoint survival setting. As a consequence the gold standard in this setting is the one minus Kaplan-Meier estimator which is compared to the incidence proportion (see Figure~\ref{fig:composite}).

In the composite endpoint analysis the underestimation of the incidence proportion is more pronounced than in the analyses of the AE probability presented above. One reason is that even in the presence of censoring for the one minus Kaplan-Meier estimator the type of the last event is most important. If the last event is an AE or CE the one minus Kaplan-Meier estimator is equal to one, even though censoring has been observed at earlier follow-up times. The incidence proportion is only equal to one if no censoring is observed. 

\section{Conclusion}
The starting point of the present investigation was that AE analyses in terms of AE probabilities, an
important aspect of drug safety evaluations, should account for the time under observation and censoring if the latter is imposed by the data at
hand. However, while primary efficacy endpoints often are time-to-event
composites such as progression-free and overall survival which every patient
experiences, although possibly after study closure, the occurrence of AE (of a
certain type) usually is subject to CEs such as death before
AE. Survival analyses accounting for CEs is methodologically well
established, but practical use lacks behind
\cite{schumacher2016competing,phillips_20}. Failure to account for censoring (e.g.,
incidence proportion) or CEs (e.g., one minus Kaplan-Meier) will
generally lead to biased quantification of absolute AE risk, but the amount of
bias has been unclear.

In this study, we confirmed that one minus Kaplan-Meier should not be used to
estimate the cumulative AE probability, as it is bound to overestimate as a
consequence of ignoring CEs. Interestingly, we found that the
incidence proportion performed surprisingly well when compared to the
gold-standard Aalen-Johansen estimator. One reason may be a high amount of
CEs before possible censoring. But not only the proportion of censoring but also the timing of the censoring are relevant as the first example of the single trials described in detail showed. This example led to the largest bias although the proportion of censoring was below average. The observed proportion and timing of censoring in this project are a consequence of twelve out of 17 trials being from oncology in which compared to other therapeutic areas AEs and CEs are often observed early during follow-up and censoring occurs much later. We also note that the observed constellation of CEs and censoring results from a sample of completed trials after the final analysis had been performed. The proportion of censoring may be different at the time point of a safety interim analysis of trials which are typically presented to data safety monitoring boards. For this situation the different estimators may behave differently \cite{hollaender2020}.

This finding must not be interpreted as a carte blanche to use AE incidence
proportions based on censored data. In fact, comparable performance of
incidence proportion and Aalen-Johansen did not only rely on a high amount of
CEs, but in particular on a careful definition of what kind of
events constitute a competing event as outlined earlier. In other words, use
of the incidence proportion \emph{implicitly} assumes events to be competing
as defined in the methods section. This aspect is somewhat subtle, but nicely
highlighted by the fact that an analysis accounting for both censoring and
only death as CEs (Aalen-Johansen (death only)) also led to
overestimating AE risk, although the bias was not as pronounced as for one
minus Kaplan-Meier.

We also found that previous worries about the constant hazard assumption
underlying incidence densities were justified in that a simple transformation
of the AE incidence density onto probabilities (probability transform incidence 
density ignoring CE) performed worst. However,
accounting for competing events in an analysis that parametrically mimicked
the non-parametric Aalen-Johansen performed better than both one minus
Kaplan-Meier and Aalen-Johansen (death only); in this sense, ignoring
CEs appeared to be worse than assuming constant hazards in our
empirical study.

Most of the results were shown for the situation where the maximum follow-up time were chosen as evaluation time.
When looking at earlier evaluation times defined by quantiles of the observed times, the resulting bias was, in general, less pronounced, 
due to a reduced relative frequency of competing events and of censoring (see figure 1). 
We regarded the situation of including all data up to the maximum follow-up time as the most relevant as this is the usual 
practice.

Our empirical study does have shortcomings. Using an opportunistic sample of
randomized clinical trials from several sponsor companies, we have been able
to illustrate possible consequences when quantifying AE risk in a manner that
ignores censoring or CEs. However, being opportunistic, the sample
does not lend itself to straightforward generalizations. More than two thirds
of the trials were from oncology. These came with a high amount of CEs, which, in turn, led to comparable performances of incidence proportion
and Aalen-Johansen. The vast majority of AEs were classified as `common' or
`very common', and AEs were also heterogeneous, coming from different
therapeutic areas and were not necessarily treatment-related. These
shortcomings were to be anticipated from an opportunistic sample, but it was our aim 
in this `real-world' setting to investigate and demonstrate which biases can occur in practice. These shortcomings do also impact the comparison of adverse event risks between treatment groups\cite{twosample}.
The observed results motivate future empirical investigations on how to
quantify AE risk with the aim of better generalizability. As a further point, it was not the aim of this investigation to accurately estimate AE probabilities, but to compare different estimators. Our present study does not allow for a meaningful comparison of results in different diseases. Follow-up investigations concentrating on trials in specific disease areas are planned.

A methodological restriction is that we have focused our investigation on an
analysis which mostly does not consider AEs after treatment discontinuation due to e.g. disease
progression in oncology. This restriction is, in particular, due to trial design when treatment 
discontinuation leads to stopping AE recording after a pre-specified time period. In addition, 
in oncology, it is not uncommon that patients enter a different clinical trial after 
progression which further complicates matters. 
Another methodological restriction is that we did not consider recurrent AEs, but only
first events. It is desirable to consider more complex event histories, also
beyond time-to-first-event. However, any such consideration will need to
account for CEs (and censoring), and our investigation therefore
also informs methodological considerations for analysing such more complex
event histories. In other words, both AEs after treatment discontinuation and recurrent AEs
will still be subject to competing events.

Our recommendation is to `play it safe' when analysing AE risk in a
time-to-first-event analysis and neither hope for a small amount nor a large
amount of CEs nor a favorable interplay of the distributions of the times of AEs, CEs, and censorings. In the former case, one minus Kaplan-Meier might
work well, while in the latter two cases the incidence proportion might do
so. Playing it safe, we recommend using the Aalen-Johansen estimator which
equals one minus Kaplan-Meier in the absence of CEs and equals the
incidence proportion in the absence of censoring and does low for presence
of both CEs and censoring. Guidelines for reporting AEs should,
therefore, advocate the Aalen-Johansen estimator instead of incidence
proportion, incidence density and one minus Kaplan-Meier.


\section*{Data and code}
Individual trial data analyses were run within the sponsor
organizations using SAS and R software provided by the
academic project group members. Only aggregated data
necessary for meta-analyses were shared and meta-analyses
were run centrally at the academic institutions.

A markdown file providing exemplary code to compute all the estimators discussed in this paper for a given dataset is available on github: \url{https://github.com/numbersman77/AEprobs}. The corresponding output is available as html file: \url{https://numbersman77.github.io/AEprobs/SAVVY_AEprobs.html}.

\section*{Funding}
Not applicable.

\section*{Declaration of conflicting interests}
KR and TK are employees of F.\ Hoffmann-La Roche (Basel, Switzerland). VJ and AB are employees of Novartis Pharma AG (Basel, Switzerland). LE, KK, FLa, FLe, AL, CN and VF are employees of Janssen-Cilag GmbH (Neuss, Germany), Bristol-Myers-Squibb GmbH \& Co. KGaA (M\"unchen, Germany), Lilly Deutschland GmbH (Bad Homburg, Germany), Pfizer Deutschland (Berlin, Germany), Merck KGaA (Darmstadt, Germany), Bayer AG (Wuppertal, Germany) and Boehringer Ingelheim Pharma GmbH \& Co. KG (Ingelheim, Germany), respectively. TF has received personal fees for consultancies (including data monitoring committees) from Bayer, Boehringer Ingelheim, Janssen, Novartis and Roche, all outside the submitted work. JB has received personal fees for consultancy from Pfizer, all outside the submitted work. CS has received personal fees for consultancies (including data monitoring committees) from Novartis and Roche, all outside the submitted work. The companies mentioned contributed data to the empirical study. RS has declared no conflict of interest.

\bibliographystyle{SageV}
\bibliography{savvy.bib}

\end{document}